\documentclass{aa}
\usepackage{epsf}
\newcommand{\ks}{\mbox{KS~1731-260 }}

\begin{document}
\sloppypar

%
   \title{Possible 38 day X-ray period of KS~1731-260}

   \author{M. Revnivtsev\inst{1,2}, R. Sunyaev \inst{2,1}}  

   \offprints{revnivtsev@hea.iki.rssi.ru}

   \institute{Space Research Institute, Russian Academy of Sciences,
              Profsoyuznaya 84/32, 117810 Moscow, Russia,
        \and
                Max-Planck-Institute f\"ur Astrophysik,
              Karl-Schwarzschild-Str. 1, 85740 Garching bei M\"unchen,
              Germany
                           }
  \date{}

        \authorrunning{Revnivtsev \& Sunyaev}
        \titlerunning{Possible 38 day X-ray period of KS~1731-260}
        
   \abstract{We report the detection of a 38 day period in the
X-ray flux of the transient burster \ks. The narrow peak of periodicity was 
detected during $\sim$TJD 10150--11050 when the source had a high and relatively 
stable X-ray flux. After $\sim$TJD 11100 the source became strongly
variable on a time scale of months that contaminates the search 
for the 38 day periodicity. The detected period can not be a binary period.
The binary with Roche lobe overflow has in this case large radii of the 
secondary and of the accretion disk. Disk and secondary star illumination 
by X-ray flux from luminous neutron star would lead to high infrared brightness
 of the binary. That clearly contradicts the infrared data even for the 
brightest infrared sources within CHANDRA error box of \ks. 
Remaining possibility is that observed periodicity is connected with the
accretion disk precession, similar to that was observed for 
SS 433, Her X-1, Cyg X-1 etc.
}
\maketitle
%

\section{Introduction}
\ks was discovered in 1989 by the TTM coded mask telescope aboard the MIR-KVANT 
observatory (\cite{discovery}, \cite{sun90}). The archived observations
of the TTM telescope showed that the source was not active more than a year before
that time
with upper limits $\sim$20 mCrab (see \cite{kolya02}).
 Since its discovery the source was extensively 
observed by various X-ray observatories. In the late 1990s, the source became
strongly variable on a month time scale, gradually decreased its X-ray 
activity and in 2001 it was found already in the
quiescent state (see \cite{wijnands01}). 

In 1997 nearly 
coherent oscillations were found in the power spectra of this source 
during a type I X-ray burst (\cite{smith97}) with a frequency $f\sim524$Hz
 (see \cite{muno00} for a review).
 It is believed that the compact object in the system \ks is one of the most
rapidly ($p\approx1/f\approx$1.9 msec) rotating neutron stars.
The distance of the source of $d\sim$7--8 kpc was estimated based on the
 analysis of the type I X-ray bursts with photospheric radius expansion 
(\cite{smith97}, \cite{muno00}).

 The source
is located close to the Galactic plane and therefore is highly reddened.
 The recent precise localization 
of the source with the Chandra observatory strongly increased the possibility 
that the real infrared counterpart of \ks will be found soon
 (see \cite{barret98}, \cite{wijnands_atel}, \cite{mikej02}).

In this Letter we present the analysis of All Sky Monitor (ASM) data
aboard the Rossi X-ray Timing Explorer (RXTE) and show the presence of 
statistically a significant periodicity in the X-ray flux of \ks.

\section{Data analysis and results}
ASM (\cite{asm}) has an intrinsic angular resolution 
of a few arcminutes, which permits to resolve bright objects even in the very 
crowded Galactic center region. The ASM provides 90-sec measurements of the 
source flux in three broad energy channels (1.5-3.0 keV, 3-5 keV and 5-12 keV)
almost every 1.5 hours. \ks was monitored by ASM since Feb. 1996.

\begin{figure}
\epsfxsize=8cm
\epsffile[53 150 430 700]{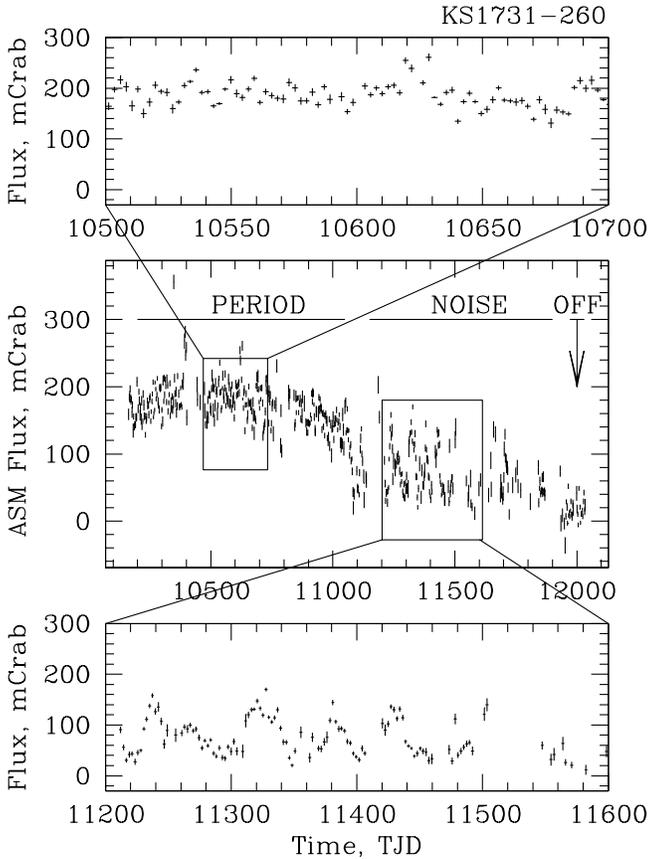}
\caption{RXTE/ASM light curve of \ks in 1996--2001. The data are binned 
into 2.5-day bins. Solid lines above the lightcurve denotes approximately 
the time periods when the periodicity and broad noise were observed. The 
upper and lower panels show more detailed view of the lightcurve of \ks during 
the periods relatively high flux and period of strong low frequency noise
 correspondingly. \label{lcurve}}
\end{figure}

The light curve of the source obtained by RXTE/ASM (1996-2001)
 is presented in Fig.\ref{lcurve} (see also \cite{wijnands01}, \cite{kolya02}).
 The flux history of \ks could be naturally subdivided into two main parts:
\begin{enumerate}
\item[a)] $\sim$TJD10150-11050 , when the source had a relatively stable and high flux
\item[b)]$\sim$TJD11100-12000 when strong variability appeared on a month time scale and
the source weakened more than twice. 
\end{enumerate}

\begin{figure}
\epsfxsize=8cm
\epsffile[53 150 460 700]{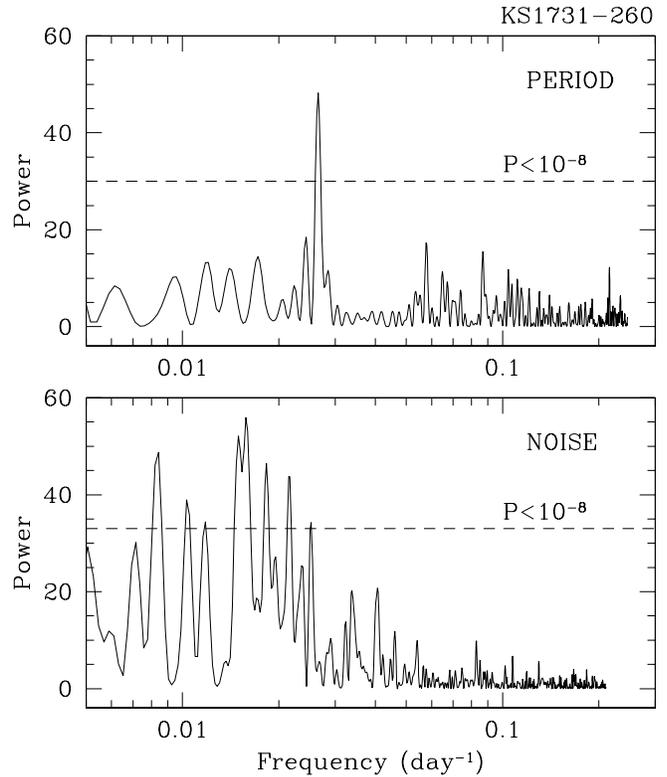}
\caption{The Lomb-Scargle periodogram of the light curve of \ks. The upper panel represents the periodogram made of points $\sim$TJD10150--11050, the lower panel - after $\sim$TJD11100--12000. The dashed line represents the $10^{-8}$ confidence level for the power values, assuming the
exponential distribution of noise powers with observed average value (the number of trial points is taken into account)\label{per_fast}}
\end{figure}

We searched for periodicities using 
the Lomb-Scargle periodograms method (\cite{lomb}, \cite{scargle}, \cite{press}) 
using the light curves of the source in the whole energy band of ASM sensitivity (1.5-12 keV).
The method of Lomb-Scarge periodograms is the most suitable for the ASM points,
because they are distributed non evenly over the total time of observations.
Due to the dependence of the real periodicity peak on the number of points
 used in the Lomb-Scargle periodogram evaluation, for this analysis we used 
lightcurves of the source with several different binning, from 
$t_{\rm bin}\sim$10 hours to 3 days. The typical Lomb-Scargle periodograms
for two abovementioned parts of the flux history of \ks  are presented 
in Fig.\ref{per_fast} (for this figure the time binning of
 the original light curves was chosen to be $\sim$6 hours). 

\begin{figure}
\epsfxsize=8cm
\epsffile[0 200 570 600]{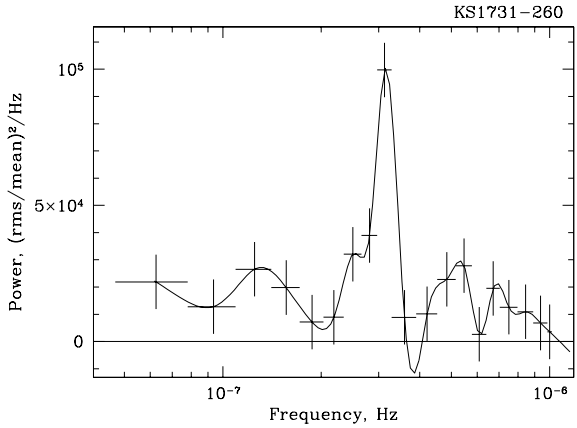}
\caption{The power spectrum of \ks according to RXTE/ASM data. Original light curve was binned in 4-days bins. The solid line is simple spline to the obtained points.\label{power}}
\end{figure}

{\em a) For the first part of the lightcurve}
 there exist a narrow, prominent and 
statistically significant 
($>6\sigma$) peak of the power spectrum of \ks at a frequency, corresponding 
to a period $\sim37.67$ days.
  
\begin{figure}
\epsfxsize=8cm
\epsffile[53 200 570 700]{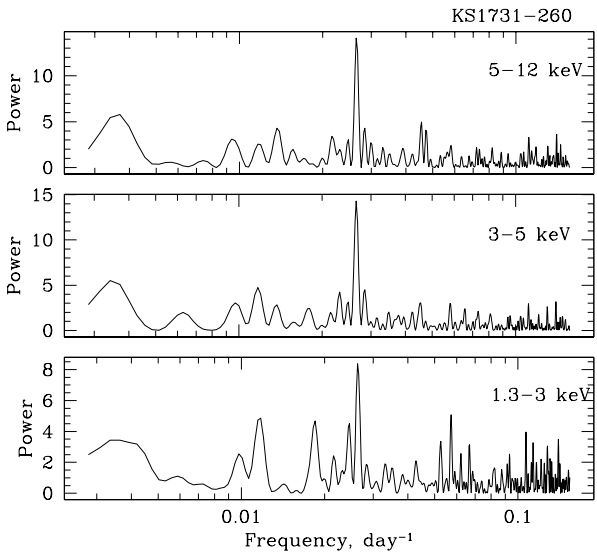}
\caption{Three Lomb-Scargle periodograms for three independent ASM 
energy channels , 1.5--3 keV, 3--5 keV and 5--12 keV.\label{three_ls}}
\end{figure}

We estimated the uncertainty of the period value 
(first part of the lightcurve, before TJD11000),
 using the Monte-Carlo bootstrap method, 
assuming the Gaussian noise of the initial light curve. The statistical 
uncertainty of the period value is $\sim$0.03 days (P$=37.67\pm0.03$ days).
In order to obtain an additional proof of the presence of the real
 periodicity in the ASM signal, we made the simple Fourier power spectrum
 of the highly binned ASM lightcurve (see Fig.\ref{power}), and also made 
the Lomb-Scargle periodograms for all three energy channels of the ASM 
instrument. The resulted power spectra are presented in Fig.\ref{three_ls}.
It is seen that the noisy peaks at the periodograms are changing, while the 
38 days period is present in all three channels. However, in the softest (1.5-3 keV)
 energy channel this peak is not statistically significant. It seems that 
the periodicity in the X-ray flux of \ks does not belong 
preferentially to some particular energy channel of ASM. 
  
\begin{figure}
\epsfxsize=8cm
\epsffile[53 200 560 560]{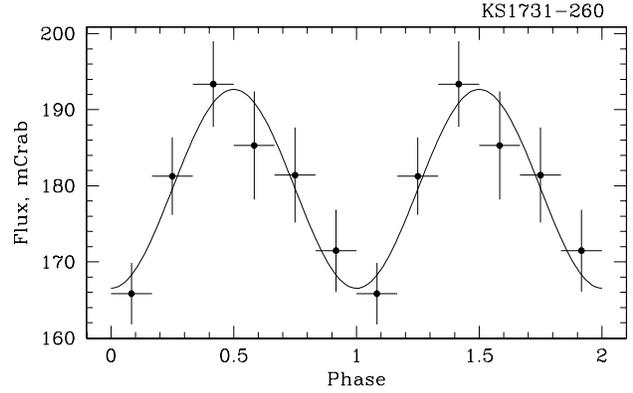}
\caption{The folded lightcurve of \ks. The error bars represents the rms deviations of the points within phase intervals (i.e. not statistical uncertainties).\label{pulse}}
\end{figure}

  Then we folded the 
light curve of \ks with the best-fit period of 37.67 days. The resulted 
phase-intensity diagram of the source is presented in Fig.\ref{pulse}.
The folded curve could be well fitted by a sinusoidal wave with amplitude
$6\pm1$\%. The amplitudes of the variations in three energy channels are 
compatible. The best fit epoch for the minimum flux is: $T_0=10149.05\pm1.02$
 TJD. The persistence and coherence of the detected variations were tested on 
the subdivided ASM lightcurves. But, as far as the time interval of our 
analysis is not very large for our period ($\sim$24 periods over $\sim$2.5 
years), we 
used only two smaller intervals, approximately 450 days each. The best fit
amplitudes and the phases of the sinusoidal waves on the folded light curve 
for each interval are compatible. 

{\em b) During the second segment of the lightcurve of \ks} 
broad, statistically significant peaks are visible at frequencies 
0.01--0.03 day$^{-1}$ (see Fig.\ref{per_fast}, lower panel). The narrow
peak at $\sim$38 day is not visible anymore. However, it should be noted
here that numerical simulations showed that the upper limit of the 
amplitude of the possible 38-day variability is not very stringent $<$10-15\%. 
This limit is sufficiently higher than the amplitude of the 38 day variations,
detected during the first period (see Fig.\ref{pulse}).

\section{Discussion}

Three types of periodicity are known in X-ray binaries: 1) the period of the neutron star
rotation, ranging from milliseconds to hundreds of seconds; 2) the orbital period
of the binary system, from tens of minutes to tens of days; and 3) the period of the
accretion disk precession, as it was observed, e.g. in the case of SS433, Her X-1, Cyg X-1 etc.

Observations of nearly 
coherent oscillations during type I X-ray bursts from \ks strongly suggest 
that the 
period of a neutron star rotation in this system is close to 
1.9 msec (\cite{smith97}, \cite{muno00}).

\subsection{Binary period?}
 Let us assume that the binary period of
\ks is $\sim$38 days and the accretion of the binary system goes due to
Roche lobe overflow. The value of the period in this case gives us the 
binary system separation: $a\sim3.69\cdot10^{12}(1+q)^{1/3}$ cm 
(assuming the mass
 of the primary $M_{NS}=1.4 M_{\odot}$), where $q$ -- is the mass ratio
 of the companions (see e.g. \cite{warner}). The star should have a radius
$R_2\sim3.69\cdot10^{12}(1+q)^{1/3}(0.38+0.2\log q)$ cm 
(assuming that $0.3<q<20$). Taking $q>0.3$, the minimal possible radius 
of the secondary could be estimated to be $R_2\ga15R_{\odot}$.

Now, using this minimal radius of the secondary
 and the source distance $d\sim$7-8 kpc we can 
calculate the minimal apparent infrared magnitudes of the binary for any
assumed temperatures of the secondary star surface. These estimates  
are very important in the view of the progress, that was achieved recently in the 
search of the \ks infrared counterpart (see e.g. \cite{barret98},
\cite{wijnands_atel}, \cite{mikej02}, Wijnands, private communication).
The brightest infrared object, that is located within a 1.5\arcsec   radius from the 
Chandra position of \ks had an apparent magnitudes $m_J\sim$16, and $m_H\sim17$ during the infrared observations of \cite{barret98}.  These observations
took place when X-ray source was luminous ($L_{\rm x}\sim4\cdot10^{37}$ erg/s). The limits $m_J\ga16$ and $m_H\ga17$ could only be satisfied
 if the $15R_{\odot}$-star surface temperature is less than $T\sim$1800K 
(the boundaries of the J and H bands and the reference fluxes were 
taken from \cite{zombeck}). The extinctions $A_J\sim2$ and $A_H\sim1.25$ 
(see \cite{barret98}) were taken 
into account when we were making these estimates. 
The obtained upper limit on the surface temperature seems 
unreasonable taking into account the significant irradiation of the 
secondary star by the neutron star X-ray emission.
 Note here, that the size of the accretion disk usually has a radius 
comparable with the radius of the star, and the solid angle of the accretion 
disk, viewed from the neutron star surface is larger than that of the 
secondary. Illumination of the accretion disk by the X-ray source
leads to its high brightness in the infrared spectral band (see
e.g. \cite{lyutyi76}) which also strongly 
contradicts the existing infrared observations.

 We can conclude that it is very unlikely that the observed 38 day
periodicity in the X-ray light curve of \ks is connected with the binary
 period of the system.

The presented consideration supports the assumption that \ks is a low mass 
X-ray binary and we can anticipate that its infrared brightness should
strongly decrease after the turnover of the X-ray source.

\subsection{Precession period?}

It is possible that the observed periodicity in the 
X-ray flux of \ks  is connected with the precession of the accretion disk, 
similarly to what is observed in the case of some other known X-ray binaries. 
Several X-ray binaries demonstrate precession of the accretion disk (e.g. Her X-1, \cite{herx1_giacconi}; 
SS433, \cite{margon84}, \cite{cherepss}; GRO J1655--40, \cite{hjellming95}) 
etc. Note however, that in two latter cases the evidence for the disk 
precession is coming mainly from the precession of the relativistic jets.

\begin{acknowledgements}
Authors thank Marat Gilfanov, Eugene Churazov, Alexey Vikhlinin for 
useful discussions and Rudy Wijnands for useful remarks about the paper and 
for the important information about 
recent infrared observations of \ks. This research has made use of data 
obtained through the High Energy Astrophysics Science Archive Research 
Center Online Service, provided by the NASA/Goddard Space Flight Center. 
\end{acknowledgements}

\end{document}